# Computational Fluid Dynamics on Quantum Computers

Madhava Syamlal, Carter Copen, Masashi Takahashi
**QubitSolve Inc.**
Morgantown, WV, USA

Benjamin Hall
**Infleqtion**
Chicago, IL, USA

## 1. Abstract

QubitSolve is working on a quantum solution for computational fluid dynamics (CFD). We have created a variational quantum CFD (VQCFD) algorithm and a 2D Software Prototype based on it. By testing the Software Prototype on a quantum simulator, we demonstrate that the partial differential equations that underlie CFD can be solved using quantum computers.

We aim to determine whether a quantum advantage can be achieved with VQCFD. To do this, we compare the performance of VQCFD with classical CFD using performance models. The quantum performance model uses data from VQCFD circuits run on quantum computers. We define a key performance parameter $Q_{5E7}$, the ratio of quantum to classical simulation time for a size relevant to industrial simulations. Given the current state of the Software Prototype and the limited computing resources available, we can only estimate an upper bound for $Q_{5E7}$. While the estimated $Q_{5E7}$ shows that the algorithm's implementation must improve significantly, we have identified several innovative techniques that could reduce it sufficiently to achieve a quantum advantage. In the next phase of development, we will develop a 3D minimum-viable product and implement those techniques.

## 2. Background

QubitSolve aims to utilize quantum computers to solve the partial differential equations underlying computational fluid dynamics (CFD). This solution will enable CFD simulations that may be too slow or impossible to run on classical computers. Such simulations will generate significant economic benefits for companies that use CFD.

CFD is an engineering tool that predicts fluid flow in or around devices by solving the Navier-Stokes equations,

$$\frac{\partial \rho}{\partial t} + \frac{\partial \rho u_j}{\partial x_j} = 0$$

$$\frac{\partial u_i}{\partial t} + u_j \frac{\partial u_i}{\partial x_j} = g_i - \frac{1}{\rho}\frac{\partial P}{\partial x_i} + \nu \left[ \frac{\partial}{\partial x_j}\left( \frac{\partial u_i}{\partial x_j} + \frac{\partial u_j}{\partial x_i} \right) - \frac{2}{3}\frac{\partial}{\partial x_i}\left( \frac{\partial u_j}{\partial x_j} \right) \right]$$

where $\rho$ is the density, $t$ is the time, $u_j$ is the velocity component in the spatial direction $x_j$, $g_i$ is the acceleration due to an external force such as gravity, $P$ is the pressure, $\nu$ is the kinematic viscosity, and repeated indices imply summation.

CFD is used in applications such as designing airplanes and automobiles, scaling up chemical reactors, and predicting ocean flow around offshore drilling platforms. CFD simulations enable engineers to efficiently explore numerous designs at low cost and arrive at optimal designs rapidly.

The speed of CFD is limited by the amount of data that must be stored in the memory of classical computers. It is not the floating-point operations but the data movement that determines the computational speed of modern supercomputers. Due to speed considerations, most industrial CFD simulations use the Reynolds-averaged Navier-Stokes (RANS) method, which is an approximate but cost-effective approach. However, RANS may not be suitable for certain industrial applications, such as simulations of an aircraft during take-off or landing [1], where more accurate large-eddy simulations (LES) are required. Such simulations could save, for instance, hundreds of millions of dollars of the $1-2 billion companies spend to certify a commercial aircraft for service [2]. According to NASA's CFD Vision 2030 Roadmap, industrial LES applications may not be available by 2030 [3].

The memory limitations of classical computers could be overcome by utilizing the vast quantum state spaces that have recently become available through quantum computers. However, quantum computers are not well-suited for solving non-linear problems, and the qubits currently available are noisy. A variational algorithm, which uses a quantum computer acting as a co-processor to a classical computer, can overcome these limitations. Much research is being conducted on variational quantum algorithms for chemistry, optimization, linear algebra, and other applications. It is a leading contender for achieving quantum advantage on noisy, intermediate-scale quantum (NISQ) computers, e.g., [4], [5], [6], [7].

Several quantum CFD methods have been documented in the literature: lattice-gas models [8], [9], [10], [11], [12]; replacing the Poisson or linear equation solver with a quantum method [13], [14], [15], [16]; solving Burgers equation [5], [17], [18]; and several others [19], [20], [21], [22], [23], [24], [25], [26], [27], [28], [29], [30], [31], [32]. QubitSolve has developed a variational quantum CFD (VQCFD) algorithm based on the lattice Boltzmann method (LBM) [33]. Our current focus is on solving the incompressible form of Navier-Stokes equations without an external force. To our knowledge, our algorithm is first-of-its kind [21], [34], [35].

This paper reports on the progress we have made in implementing this algorithm in a Software Prototype and assessing its performance on quantum computers.

## 3. Software Prototype Development

VQCFD encodes CFD data as the amplitudes of quantum state spaces. It ensures that CFD data of size $\mathcal{O}(N_{grid})$, where $N_{grid}$ is the number of grid points, lives only on the quantum computer (see Figure 1). VQCFD achieves this by using parametrized quantum circuits (PQC) to represent the CFD fields. The PQCs generate the CFD data using input parameters $\boldsymbol{\theta}$, vectors of size $\mathcal{O}\left(\log\left(N_{grid}\right)\right)$. The classical computer only needs to store and handle $\boldsymbol{\theta}$ to evolve the CFD solution on the quantum computer, which is the potential source of quantum advantage. This advantage is gained by trading off the knowledge of the complete solution at any timestep on the classical computer. However, the complete solution is often unnecessary because CFD analyses usually seek a subset of the solution that can be efficiently extracted from the PQCs, e.g., integral quantities such as the lift and drag coefficients in an aircraft simulation [28].

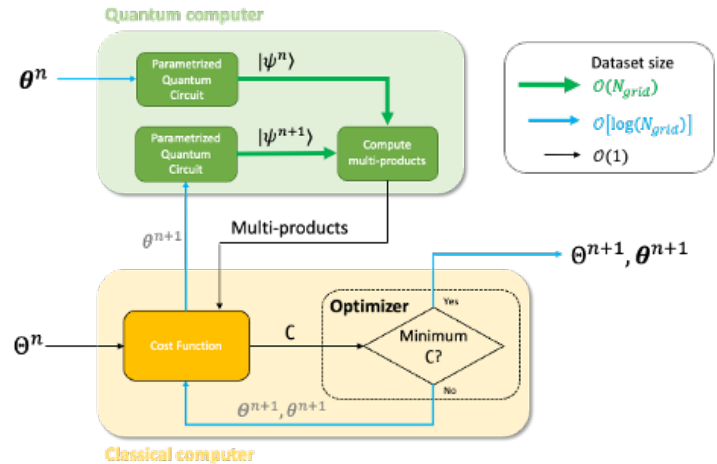

*Figure 1. The VQCFD algorithm ensures that CFD fields of size $\mathcal{O}\left(N_{grid}\right)$ resides only on the quantum computer. These fields are generated by PQCs using input parameters $\boldsymbol{\theta}$ of size $\mathcal{O}\left(\log\left(N_{grid}\right)\right)$. The fields are encoded into the amplitudes of the quantum state $|\psi\rangle$. The quantum computer calculates multi-products, the sum of the products of the corresponding elements of multiple vectors. The multi-products are used by the classical computer to compute a cost function. The minimum value of the cost function corresponds to the solution of the CFD equations.*

In a project funded by the National Science Foundation we implemented a 2D version of VQCFD in a Software Prototype. The prototype consists of a classical CFD code and a VQCFD code, both based on the D2Q9 velocity set of LBM [33]. The software is written in Python and uses Qiskit, an open-source software development kit for programming quantum computers. The classical CFD code handles periodic, mass-inflow, pressure-outflow, and moving-wall boundary conditions. The VQCFD code handles periodic and moving-wall boundary conditions. Both codes use an input data file to set up a CFD simulation.

To ensure the accuracy of the Software Prototype, we perform unit testing and solution verification. We use several unit tests to verify the Software Prototype methods systematically. For example, to test a method called 'train_PQC,' we use a random array as input. Then, we determine the PQC parameters using 'train_PQC.' We use those parameters to create a quantum circuit from which the field variable values are recovered. We compare the recovered values with the input array to verify the PQC training method.

We perform solution verification by comparing the VQCFD solution with the classical CFD solution. VQCFD is much slower than classical CFD for small simulations. Thus, we can repeat calculations with classical CFD at a negligible increase in the computational cost. We verify the solution in two ways: by solving VQCFD equations algebraically and variationally. First, we compare the algebraic solutions of classical CFD and VQCFD to verify the VQCFD method for implementing boundary conditions. We verify that the two solutions agree with machine precision.

Second, we calculate the classical CFD solution at each time step and determine the exact minimum value of the cost function that the variational method must achieve to reach that solution. Then, we verify that VQCFD iterations converge toward the exact minimum value. This verification is carried out on a quantum simulator, as the correctness of VQCFD implementation is independent of whether the circuits are run on a quantum simulator or a quantum computer. Figure 2 shows examples of the convergence behavior (for variables $v = 3$ and 6) that converged in 1000 optimizer iterations. This result is a significant milestone in our research, as it demonstrates that Navier-Stokes equations can indeed be solved using VQCFD. To the best of our knowledge, it is the first time such a demonstration has been achieved (e.g., [21], [34], [35]).

## 4. Quantum and Classical Performance Models

Having shown that it is possible to solve CFD equations leveraging quantum computers, the next question is whether there is an advantage in doing so. To answer that question, we must compare the performance of VQCFD with that of classical CFD. We compare them using the performance models described next.

### VQCFD Performance Model

Considering the wait time on the quantum job queue, the cost of quantum computing, and the speed of the interconnects between classical and quantum computers; it is currently not feasible to run the Software Prototype on a hybrid classical-quantum computer. Therefore, we developed a method to extract quantum circuits from the Software Prototype and measure their performance on quantum computers.

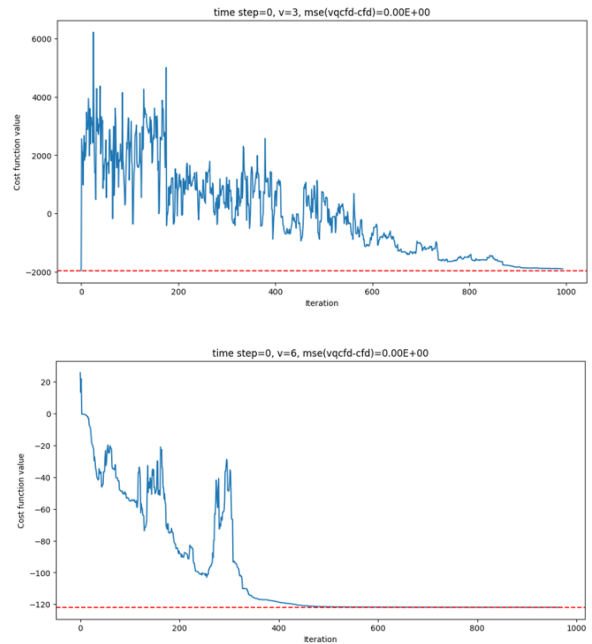

*Figure 2. The plots show how the optimizer reduces cost function values for two variables, for example. The red-dashed lines depict the exact minimum value calculated with the classical CFD code, to which the iterations converge with high accuracy. Such convergence demonstrates that the partial differential equations underlying CFD can indeed be solved on quantum computers.*

The total runtime consists of the time needed (1) to initialize the simulation, (2) to conduct the transient simulation, and (3) to finalize the simulation. The time for steps (1) and (3) is negligible compared with that for step (2) and disregarded in our model.

The runtime per time-step for the transient simulation consists of the time needed for (1) classical calculations such as optimization, (2) transferring $\boldsymbol{\theta}$ from the classical computer to the quantum computer, (3) quantum computations, and (4) transferring multi-products from the quantum computer to the classical computer (see Figure 1). The time needed for step (1) is relatively negligible. Fast enough interconnects between quantum and classical computers will likely be available, and the amount of data transferred will be small, $\mathcal{O}\big(\log\big(N_{grid}\big)\big)$ and $\mathcal{O}(1)$. So, we can ignore the time needed for steps (2) and (4).

Thus, we get the following equation for the runtime per time-step

$$t_q = \sum_{v}^{N_v} \left( N_{i,v} \, N_f \sum_{p}^{N_p} N_{shot} \, t_{p,v} \right)$$

where $t_q$ is the runtime per time-step, $N_v$ is the number of variables (= 9 for D2Q9), $N_{i,v}$ is the number of optimizer iterations for variable $v$, $N_f$ is the number of function evaluations per iteration, $N_{shot}$ is the number of shots/iteration (= 10,000), $N_p$ is the number of multi-product circuits (= 54 for VQCFD algorithm), $t_{p,v}$ is the run time of circuit $p$ in the calculations for variable $v$. $N_f$ is 2 for SPSA, the optimizer used in this study and does not vary with the number of parameters [36].

We assume that the variations with respect to $v$ are small. Furthermore, we categorize the circuits into small and large circuits. Then,

$$t_q \approx N_v \, \overline{N}_i \, N_f \, N_{shot} \left( N_{ps} \, \overline{t}_{ps} \, + \, N_{pl} \, \overline{t}_{pl} \right)$$

where $\overline{N}_i$ is the mean number of iterations, $N_{ps}$ is the number of small circuits (=9), $N_{pl}$ is the number of large circuits (=45), $\overline{t}_{ps}$ is the mean small circuit execution time, and $\overline{t}_{pl}$ is the mean large circuit execution time.

The number of optimizer iterations $\overline{N}_i$ is likely to be proportional to the number of parameters, which scales with $N_q$, the number of qubits per variable. For example, the number of parameters is $2N_q$ for the PQC used in this study. We determined $\overline{N}_i = 125N_q$ from quantum simulator runs. However, the convergence of variational quantum algorithms is still an open question [7]. It could be affected by vanishing gradients, known as the barren plateau problem [37].

The run time is proportional to the circuit depth [38], which also scales with $N_q$ [10], [39]. So, we use regression models to determine $\overline{t}_{ps}$ and $\overline{t}_{pl}$.

We've meticulously collected data for $\overline{t}_p$ on two types of quantum computers: IonQ, a trapped ion quantum computer accessed through QLab at the University of Maryland, and IBM, superconducting qubit quantum computers accessed through the Oak Ridge Leadership Computing Facility (OLCF). This comprehensive data collection process ensures the reliability of our findings.

We ran a random sample of 100 circuits for a grid size 16 ($N_q = 4$), about 21% of the VQCFD circuits, on an IonQ Aria-1 computer and determined

$$\overline{t}_{ps} = \, c_0 \, N_q$$

where $c_0 = 9.66\text{E-}3$ seconds with the 95% confidence interval (9.36E-3, 9.96E-3).

We ran a random sample of 50 circuits for a grid size of 16 on IBM quantum computers. We observed a consistent runtime value across the runs. Then, we ran one small circuit at increasing grid size and $N_q$ values (column 2 and 3 of Table 1). Our algorithm determines the qubits needed for various VQCFD circuits, and the values for the small circuits are given in column 4. The quantum computers used for the runs varied; the job scheduler selected the least busy ones (column 5). These computers have 127-qubit Eagle processors that perform 5000 hardware-aware circuit-layer operations per second (CLOPS_h). Runs 10 and 11 could be done only for small circuits; the large circuits require more than the 127 qubits available on those machines. We could not go beyond run 11 even for small circuits because the Software Prototype quit while processing the input data on our Mac mini desktop, possibly because of memory limitations. The completed job provided the circuit depth and the runtime (columns 6 and 7). Although the circuit depth scales with $N_q$, runs 6 and 10 show an anomalous increase. The following fit for small circuits has an adjusted $R^2 = 0.55$:

$$\bar{t}_{ps} = 0.0006784 + 0.00017502 \, N_q$$

*Table 1. Runtime for VQCFD circuits run on IBM quantum computers (small circuits)*

| Run No | Grid size | $N_q$ | Qubits/ circuit | Quantum computer | Circuit depth | $\bar{t}_p \times 10^4$ seconds |
|---|---|---|---|---|---|---|
| 1 | 1.60E+01 | 4 | 16 | ibm_kyoto | 249 | 5 |
| 2 | 6.55E+04 | 16 | 48 | ibm_sherbrooke | 462 | 6 |
| 3 | 2.62E+05 | 18 | 54 | ibm_sherbrooke | 617 | 6 |
| 4 | 1.05E+06 | 20 | 60 | ibm_kyoto | 759 | 9 |
| 5 | 4.19E+06 | 22 | 66 | ibm_kyoto | 808 | 7 |
| 6 | 1.68E+07 | 24 | 72 | ibm_kyoto | 1119 | 7 |
| 7 | 6.71E+07 | 26 | 78 | ibm_brisbane | 926 | 7 |
| 8 | 2.68E+08 | 28 | 84 | ibm_sherbrooke | 932 | 7 |
| 9 | 1.07E+09 | 30 | 90 | ibm_osaka | 1236 | 8 |
| 10 | 4.29E+09 | 32 | 96 | ibm_kyoto | 1744 | 9 |
| 11 | 1.72E+10 | 34 | 102 | ibm_kyoto | 1349 | 8 |

Then, we ran one large circuit at increasing grid size and $N_q$ values (column 2 and 3 of Table 2). The qubits needed for the large circuits is given in column 4. The completed job provided the circuit depth and the runtime (columns 6 and 7).

*Table 2. Runtime for VQCFD circuits run on IBM quantum computers (large circuits)*

| Run No | Grid size | $N_q$ | Qubits/ circuit | Quantum computer | Circuit depth | $\bar{t}_p \times 10^4$ seconds |
|---|---|---|---|---|---|---|
| 1 | 1.60E+01 | 4 | 16 | ibm_osaka | 231 | 5 |
| 2 | 6.55E+04 | 16 | 64 | ibm_osaka | 4817 | 15 |
| 3 | 2.62E+05 | 18 | 72 | ibm_osaka | 15535 | 36 |

| 4 | 1.05E+06 | 20 | 80 | ibm_osaka | 19793 | 44 |
| 5 | 4.19E+06 | 22 | 88 | ibm_kyoto | 29895 | 90 |
| 6 | 1.68E+07 | 24 | 96 | ibm_osaka | 112910 | 234 |
| 7 | 6.71E+07 | 26 | 104 | ibm_kyoto | 30096 | 300 |
| 8 | 2.68E+08 | 28 | 112 | ibm_kyoto | 26706 | 300 |
| 9 | 1.07E+09 | 30 | 120 | ibm_kyoto | 18538 | 300 |

The following fit for large circuits has an adjusted $R^2 = 0.86$:

$$\bar{t}_{pl} = 0.00181582 + 0.01748484\, N_q + 0.01597166 N_q^2$$

## Classical CFD Performance Model

As in quantum advantage demonstrations (e.g., [40], [41]), we use the world's fastest supercomputer as the classical computer for the comparison: Frontier [42]. It is also ideal for comparison because over 99% of its performance comes from GPUs [43], a computing platform increasingly used to speed up CFD codes.

Figure 3 shows the Frontier GPU architecture. The GPUs assign blocks of 1024 threads to compute units (CU). The threads in a block are scheduled in units of 64 threads, called wavefronts. Each wavefront is assigned to one single instruction multiple data (SIMD) unit out of the 4 SIMD units on each CU. The entire wavefront executes the instruction in eight cycles, attaining a speed of 23.9 TFLOP/s per GPU [42]. The parallel execution of the wavefronts speeds up computations.

In our model the computations for each grid point will occur on a thread. The run time per time step for classical CFD is the sum of the run times for the two LBM steps

$$t_c = t_{collision} + t_{streaming}$$

The collision step requires only local data. So, we assume all the wavefronts on all the CUs will execute in parallel. Therefore, the total run time is the run time of one wavefront multiplied by the number of wavefronts ($N_{wave}$).

The code for the collision step is divided into computational kernels, as suggested by Kothapalli et al. for their GPU execution model [44]. The run time for each kernel is estimated using the roofline model – the maximum of the time to access memory and to perform floating point operations [42], [45]. Thus,

$$t_{collision} = N_{wave} \left( \sum_{k=0}^{2} \max\left( \frac{B_k\, N_{T/GPU}}{BW_{HBM}},\ \frac{F_k}{FLOP_{max}} \right) + N_v \sum_{k=3}^{4} \max\left( \frac{B_k\, N_{T/GPU}}{BW_{HBM}},\ \frac{F_k}{FLOP_{max}} \right) \right)$$

where $B_k$ and $F_k$ are the bytes moved and floating-point operations required in the k<sup>th</sup> computational kernel, $N_{T/GPU}$ is the number of threads on a GPU, $BW_{HBM}$ is the bandwidth of the high bandwidth memory (HBM) on the GPUs, and $FLOP_{max}$ is the maximum value of floating-point operations per second. For the computational kernels in our CFD model, $F_k/B_k$ is [0.1, 0.193, 0.193, 0.425, 0.0] FLOP/byte. This ratio, called Arithmetic Intensity, is low, implying that the memory access time dominates.

The kernel corresponding to the streaming step must also account for the time for communications between GPUs on a node or different nodes, for which we use a model like [46]. The streaming step moves one double precision floating point number at every grid point to a neighboring one for all variables, except $v = 0$. That involves moving two numbers in an out of HBM for each thread on a GPU, moving $N_{bdr}$ numbers between GPUs on a node and $N_{bdr}$ numbers between nodes, where we assume the number of boundary points to be $N_{bdr} = \sqrt{N_{T/GPU}}$ (for 2D). The interconnects are said to be of low latency, and no values are reported; we assume the latencies are zero. We assume that the communications can perfectly overlap, making time for streaming equal to that for the slowest step:

$$t_{streaming} = (N_v - 1) \max \begin{pmatrix} \dfrac{2\, B_{DP}\, N_{T/GPU}}{BW_{HBM}}, \\ \dfrac{7\, B_{DP} N_{bdr}}{BW_{GPU}}, \\ \dfrac{N_{Node} B_{DP} N_{bdr}}{BW_{Node}} \end{pmatrix}$$

where $B_{DP}$ is the number of bytes per double precision floating point number, $BW_{GPU}$ is the GPU-to-GPU bandwidth, $N_{Node}$ is the total number of nodes, and $BW_{Node}$ is the inter-node bandwidth.

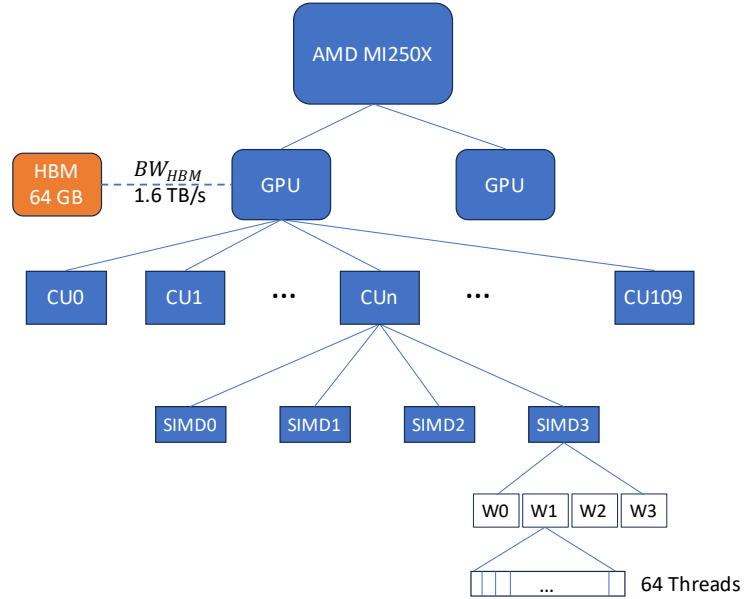

*Figure 3. The supercomputer Frontier has 9408 nodes, each connected to four AMD MI250X Accelerator Modules. Each module consists of two GPUs. Each GPU has 64 GB of high bandwidth memory (HBM) and 110 compute units (CUs), each CU containing four single instruction, multiple data (SIMD) units that execute wavefronts of 64 threads each.*

The SIMD units on the GPUs enable parallel calculation of many grid points. So, distributing the problem over Frontier's 75,264 GPUs could reduce computational time. However, as the number of GPUs increases, so does the time for communications between GPUs, which slows the streaming step. Therefore, the minimum run time for a given grid size occurs at an optimal number of GPUs, as seen in the example in Figure 4. We determine the optimum number of GPUs for each grid size and use the corresponding run time as the classical CFD time.

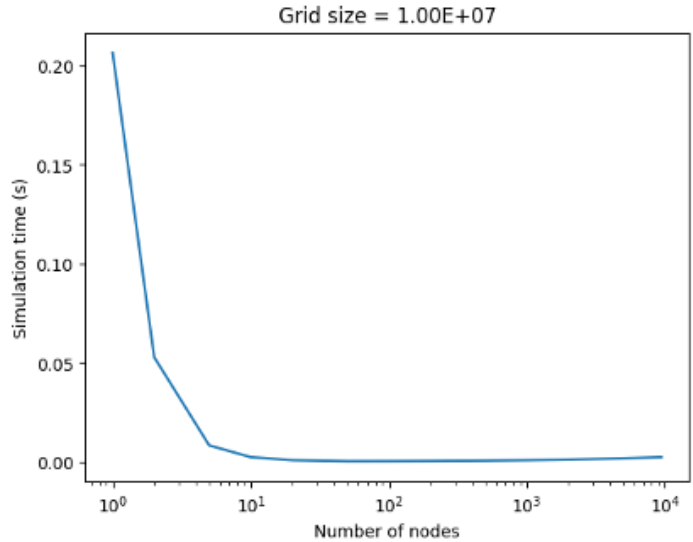

*Figure 4. The simulation time varies with the number of nodes. Initially, it decreases as more GPUs are used for computations. However, the time increases beyond an optimal number of nodes due to the time needed for inter-node communication. In the example shown, the optimal number of nodes is 52 for a grid size of ten million.*

To validate the classical performance model, we compare our predictions with the measured performance of D2Q9 run on AMD MI100 [47]. For 10 million grid points, our model predicts 38,800 million lattice updates per second (MLUPS), and the measured performance is 6,000 MLUPS. We expect the performance on Frontier to be better because its GPUs are faster than MI100, and there are more of them. Our model likely over-predicts the performance because of the assumed highly optimized implementation of the classical algorithm. To minimize the discrepancy between predicted and measured MLUPS, we selected the lowest value for $BW_{Node}$ from the reported range of 3 to 17.5 GB/s [43].

## Comparison of VQCFD and Classical CFD Performance

Figure 5 compares the predictions of performance models for classical CFD running on Frontier and VQCFD running on IBM quantum computers. VQCFD performance predicted using $\bar{t}_p$ data (Table 1 and Table 2) are shown as filled circles.

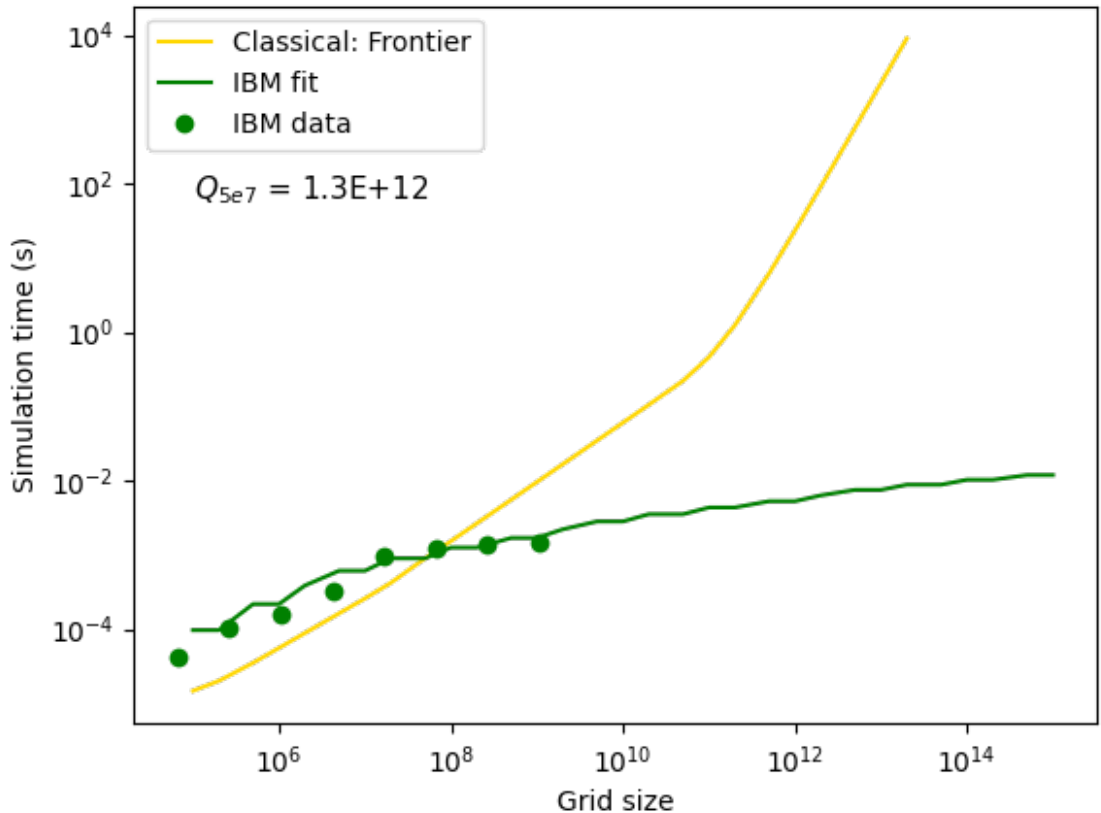

*Figure 5. The chart compares the simulation time of classical CFD with VQCFD, as their respective performance models predicted. The classical CFD performance model assumes optimal performance of our CFD code on Frontier, the world's fastest classical computer. The filled circles on the VQCFD performance model represent circuit run time data obtained from IBM quantum computers, and the line fits through these data points. The VQCFD performance is scaled by $Q_{5E7}$, the ratio of quantum to classical simulation time for a grid size of fifty million, a typical grid size used in industrial CFD simulations. The current estimate of $Q_{5E7}$ is $\mathcal{O}(10^{12})$. It is only an upper bound and is necessarily very pessimistic, as we are comparing the performance of an unoptimized quantum software prototype to an optimized implementation of classical CFD. However, we have identified innovative techniques that could reduce $Q_{5E7}$ to less than 1, achieving a quantum advantage.*

We introduce the key performance parameter $Q_{5E7}$, the ratio of quantum to classical simulation time for a grid size of fifty million, a typical grid size used in industrial CFD simulations. If $Q_{5E7} \leq 1$ a quantum advantage is achieved for grid sizes greater than fifty million. Our data indicate that the upper limit of $Q_{5E7}$ is of $\mathcal{O}(10^{12})$. This preliminary estimate is necessarily very pessimistic as we are comparing an unoptimized Software Prototype to an optimized simulation on the world's fastest classical computer. To reduce the prototype development time, we did not optimize the Software Prototype. Instead, we prioritized the verification of the algorithm and its transparent implementation. We will do the optimizations in a 3D minimum viable product that will be developed in the next phase of our development. Additionally, this value is an upper limit, and the required improvement could be smaller.

While our estimate of the required improvement appears daunting at first sight, it has allowed us to identify a series of innovative techniques – developing efficient parameterized quantum circuits, parallelizing the field variable and multi-product computations, introducing error mitigation techniques and optimized shot allocation, minimizing the depth and gate count of our circuits using Superstaq [48], using circuit decomposition based on the target quantum computer's qubit connectivity, using noise-aware mapping/routing techniques, and minimizing qubit measuring requirements – that could reduce $Q_{5E7}$ to less than 1, achieving a quantum advantage. Our future work will be centered around pursuing that target.

## 5. Conclusions

In this work, we implemented a quantum algorithm for CFD called VQCFD in a Software Prototype. With the help of the Software Prototype, we achieved two milestones:

1. We have demonstrated that quantum computers can solve the partial differential equations that underlie CFD, marking a new era of innovation in this field.
2. We have made an unprecedented, and therefore significant, comparison of the performance of classical CFD and VQCFD.

We estimate an upper bound for the key performance parameter $Q_{5E7}$, the ratio of quantum to classical simulation time for a grid size of fifty million, as $\mathcal{O}(10^{12})$. While the required improvement appears daunting at first sight, we have identified a series of innovative techniques that could reduce $Q_{5E7}$ to less than 1, achieving a quantum advantage.

**Acknowledgments**

This material is based upon work supported by the National Science Foundation under Grant No. 2318334. This research used resources of QLab, the National Quantum Laboratory at Maryland. This research used resources of the Oak Ridge Leadership Computing Facility, which is a DOE Office of Science User Facility supported under Contract DE- AC05-00OR22725.

## 6. References

[1] J. P. Slotnick and D. Mavriplis, "A Grand Challenge for the Advancement of Numerical Prediction of High Lift Aerodynamics," in *AIAA Scitech 2021 Forum*, VIRTUAL EVENT: American Institute of Aeronautics and Astronautics, Jan. 2021. doi: 10.2514/6.2021-0955.

[2] E. Nielsen, "The Impact of Emerging Architectures on Engineering and Leadership-Class Computational Fluid Dynamics," presented at the [BIG] COMPUTE 22, Nov. 09, 2022. Accessed: Jan. 27, 2023. [Online]. Available: https://player.vimeo.com/video/768734990?h=cec974686d

[3] A. Cary *et al.*, "THE CFD VISION 2030 ROADMAP: 2020 STATUS, PROGRESS AND CHALLENGES." [Online]. Available: https://www.cfd2030.com/report.html

[4] J. R. McClean, J. Romero, R. Babbush, and A. Aspuru-Guzik, "The theory of variational hybrid quantum-classical algorithms," *New J. Phys.*, vol. 18, no. 2, p. 023023, Feb. 2016, doi: 10.1088/1367-2630/18/2/023023.

[5]  M. Lubasch, J. Joo, P. Moinier, M. Kiffner, and D. Jaksch, "Variational quantum algorithms for nonlinear problems," *Phys. Rev. A*, vol. 101, no. 1, p. 010301, Jan. 2020, doi: 10.1103/PhysRevA.101.010301.

[6]  M. Cerezo *et al.*, "Variational Quantum Algorithms," *Nat Rev Phys*, vol. 3, no. 9, pp. 625–644, Aug. 2021, doi: 10.1038/s42254-021-00348-9.

[7]  J. Tilly *et al.*, "The Variational Quantum Eigensolver: a review of methods and best practices," *Physics Reports*, vol. 986, pp. 1–128, Nov. 2022, doi: 10.1016/j.physrep.2022.08.003.

[8]  J. Yepez, "Quantum Computation of Fluid Dynamics," in *Quantum Computing and Quantum Communications Lecture Notes in Computer Science*, vol. 1509, C. P. Williams, Ed., Berlin: Springer-Verlag, 1999, pp. 1–28.

[9]  J. Yepez, "Quantum lattice-gas model for computational fluid dynamics," *Phys. Rev. E*, vol. 63, no. 4, p. 046702, Mar. 2001, doi: 10.1103/PhysRevE.63.046702.

[10]  B. N. Todorova and R. Steijl, "Quantum algorithm for the collisionless Boltzmann equation," *Journal of Computational Physics*, vol. 409, p. 109347, May 2020, doi: 10.1016/j.jcp.2020.109347.

[11]  M. A. Schalkers and M. Möller, "Efficient and fail-safe collisionless quantum Boltzmann method." arXiv, Nov. 25, 2022. Accessed: Feb. 06, 2023. [Online]. Available: http://arxiv.org/abs/2211.14269

[12]  A. D. B. Zamora, L. Budinski, O. Niemimäki, and V. Lahtinen, "Efficient Quantum Lattice Gas Automata." arXiv, Feb. 26, 2024. Accessed: Jun. 20, 2024. [Online]. Available: http://arxiv.org/abs/2402.16488

[13]  Y. Cao, A. Papageorgiou, I. Petras, J. Traub, and S. Kais, "Quantum algorithm and circuit design solving the Poisson equation," *New J. Phys.*, vol. 15, no. 1, p. 013021, Jan. 2013, doi: 10.1088/1367-2630/15/1/013021.

[14]  R. Steijl and G. N. Barakos, "Parallel evaluation of quantum algorithms for computational fluid dynamics," *Computers & Fluids*, vol. 173, pp. 22–28, Sep. 2018, doi: 10.1016/j.compfluid.2018.03.080.

[15]  R. Morrison, "Rolls-Royce takes hybrid approach to quantum computing," Tech Monitor. Accessed: Dec. 20, 2022. [Online]. Available: https://techmonitor.ai/technology/emerging-technology/rolls-royce-quantum-computing-classiq

[16]  L. Lapworth, "A Hybrid Quantum-Classical CFD Methodology with Benchmark HHL Solutions," 2022, doi: 10.48550/ARXIV.2206.00419.

[17]  J. Yepez, "Quantum Lattice-Gas Model for the Burgers Equation," *Journal of Statistical Physics*, vol. 107, no. 1/2, pp. 203–224, 2002.

[18]  F. Oz, R. K. S. S. Vuppala, K. Kara, and F. Gaitan, "Solving Burgers' equation with quantum computing," *Quantum Inf Process*, vol. 21, no. 1, p. 30, Jan. 2022, doi: 10.1007/s11128-021-03391-8.

[19]  G. Xu, A. J. Daley, P. Givi, and R. D. Somma, "Turbulent Mixing Simulation via a Quantum Algorithm," *AIAA Journal*, vol. 56, no. 2, pp. 687–699, Feb. 2018, doi: 10.2514/1.J055896.

[20]  G. Xu, A. J. Daley, P. Givi, and R. D. Somma, "Quantum algorithm for the computation of the reactant conversion rate in homogeneous turbulence," *Combustion Theory and Modelling*, vol. 23, no. 6, pp. 1090–1104, Nov. 2019, doi: 10.1080/13647830.2019.1626025.

[21] P. Givi, A. J. Daley, D. Mavriplis, and M. Malik, "Quantum Speedup for Aeroscience and Engineering," *AIAA Journal*, vol. 58, no. 8, pp. 3715–3727, Aug. 2020, doi: 10.2514/1.J059183.

[22] J. A. Scoville, "Type II Quantum Computing Algorithm For Computational Fluid Dynamics," Air Force Institute of Technology, 2006.

[23] L. Budinski, "Quantum algorithm for the advection–diffusion equation simulated with the lattice Boltzmann method," *Quantum Inf Process*, vol. 20, no. 2, p. 57, Feb. 2021, doi: 10.1007/s11128-021-02996-3.

[24] L. Budinski, "Quantum algorithm for the Navier Stokes equations by using the streamfunction vorticity formulation and the lattice Boltzmann method." arXiv, Mar. 15, 2022. Accessed: Dec. 12, 2022. [Online]. Available: http://arxiv.org/abs/2103.03804

[25] F. Gaitan, "Finding flows of a Navier–Stokes fluid through quantum computing," *npj Quantum Inf*, vol. 6, no. 1, p. 61, Jul. 2020, doi: 10.1038/s41534-020-00291-0.

[26] S. S. Bharadwaj and K. R. Sreenivasan, "Quantum Computation of Fluid Dynamics." arXiv, Jul. 17, 2020. Accessed: Dec. 12, 2022. [Online]. Available: http://arxiv.org/abs/2007.09147

[27] N. Ray, T. Banerjee, B. Nadiga, and S. Karra, "Towards Solving the Navier-Stokes Equation on Quantum Computers." arXiv, Apr. 16, 2019. Accessed: Dec. 12, 2022. [Online]. Available: http://arxiv.org/abs/1904.09033

[28] K. P. Griffin, S. S. Jain, T. J. Flint, and W. H. R. Chan, "Investigation of quantum algorithms for direct numerical simulation of the Navier-Stokes equations," *Center for Turbulence Research Annual Research Briefs*, pp. 347–363, 2019.

[29] R. Demirdjian, D. Gunlycke, C. A. Reynolds, J. D. Doyle, and S. Tafur, "Variational quantum solutions to the advection–diffusion equation for applications in fluid dynamics," *Quantum Inf Process*, vol. 21, no. 9, p. 322, Sep. 2022, doi: 10.1007/s11128-022-03667-7.

[30] Z.-Y. Chen *et al.*, "Quantum Finite Volume Method for Computational Fluid Dynamics with Classical Input and Output." arXiv, Feb. 06, 2021. Accessed: Feb. 06, 2023. [Online]. Available: http://arxiv.org/abs/2102.03557

[31] C. A. Williams, A. A. Gentile, V. E. Elfving, D. Berger, and O. Kyriienko, "Quantum Iterative Methods for Solving Differential Equations with Application to Computational Fluid Dynamics." arXiv, Apr. 12, 2024. Accessed: Jun. 20, 2024. [Online]. Available: http://arxiv.org/abs/2404.08605

[32] P. Brearley and S. Laizet, "Quantum Algorithm for Solving the Advection Equation using Hamiltonian Simulation." arXiv, Apr. 25, 2024. Accessed: Jun. 20, 2024. [Online]. Available: http://arxiv.org/abs/2312.09784

[33] T. Krüger, H. Kusumaatmaja, A. Kuzmin, O. Shardt, G. Silva, and E. M. Viggen, *The Lattice Boltzmann Method: Principles and Practice*, 1st ed. 2017 edition. New York, NY: Springer, 2016.

[34] H. P. Paudel *et al.*, "Quantum Computing and Simulations for Energy Applications: Review and Perspective," *ACS Eng. Au*, vol. 2, no. 3, pp. 151–196, Jun. 2022, doi: 10.1021/acsengineeringau.1c00033.

[35] S. Succi, W. Itani, K. Sreenivasan, and R. Steijl, "Quantum computing for fluids: Where do we stand?," *EPL*, vol. 144, no. 1, p. 10001, Oct. 2023, doi: 10.1209/0295-5075/acfdc7.

[36] A. Szava and D. Wierichs, "Optimization using SPSA," *PennyLane Demos*, Mar. 2023, Accessed: May 07, 2024. [Online]. Available: https://pennylane.ai/qml/demos/tutorial_spsa/

[37] J. R. McClean, S. Boixo, V. N. Smelyanskiy, R. Babbush, and H. Neven, "Barren plateaus in quantum neural network training landscapes," *Nat Commun*, vol. 9, no. 1, p. 4812, Nov. 2018, doi: 10.1038/s41467-018-07090-4.

[38] N. S. Blunt *et al.*, "Perspective on the Current State-of-the-Art of Quantum Computing for Drug Discovery Applications," *J. Chem. Theory Comput.*, vol. 18, no. 12, pp. 7001–7023, Dec. 2022, doi: 10.1021/acs.jctc.2c00574.

[39] S. Sim, P. D. Johnson, and A. Aspuru-Guzik, "Expressibility and entangling capability of parameterized quantum circuits for hybrid quantum-classical algorithms," *Adv Quantum Tech*, vol. 2, no. 12, p. 1900070, Dec. 2019, doi: 10.1002/qute.201900070.

[40] F. Arute *et al.*, "Quantum supremacy using a programmable superconducting processor," *Nature*, vol. 574, no. 7779, pp. 505–510, Oct. 2019, doi: 10.1038/s41586-019-1666-5.

[41] Y.-H. Deng *et al.*, "Gaussian Boson Sampling with Pseudo-Photon-Number Resolving Detectors and Quantum Computational Advantage." arXiv, Sep. 01, 2023. Accessed: May 04, 2024. [Online]. Available: http://arxiv.org/abs/2304.12240

[42] "Frontier User Guide — OLCF User Documentation." Accessed: May 06, 2024. [Online]. Available: https://docs.olcf.ornl.gov/systems/frontier_user_guide.html

[43] S. Atchley *et al.*, "Frontier: Exploring Exascale," in *Proceedings of the International Conference for High Performance Computing, Networking, Storage and Analysis*, Denver CO USA: ACM, Nov. 2023, pp. 1–16. doi: 10.1145/3581784.3607089.

[44] K. Kothapalli, R. Mukherjee, M. S. Rehman, S. Patidar, P. J. Narayanan, and K. Srinathan, "A performance prediction model for the CUDA GPGPU platform," in *2009 International Conference on High Performance Computing (HiPC)*, Kochi, India: IEEE, Dec. 2009, pp. 463–472. doi: 10.1109/HIPC.2009.5433179.

[45] X. Lu and N. Wolfe, "Hierarchical Roofline on AMD Instinct$^{TM}$ MI200 GPUs," presented at the Supercomputing 2023, Denver, CO, Nov. 2023.

[46] B. V. Werkhoven, J. Maassen, F. J. Seinstra, and H. E. Bal, "Performance Models for CPU-GPU Data Transfers," in *2014 14th IEEE/ACM International Symposium on Cluster, Cloud and Grid Computing*, Chicago, IL, USA: IEEE, May 2014, pp. 11–20. doi: 10.1109/CCGrid.2014.16.

[47] P. Valero-Lara, J. Vetter, J. Gounley, and A. Randles, "Moment Representation of Regularized Lattice Boltzmann Methods on NVIDIA and AMD GPUs," in *Proceedings of the SC '23 Workshops of The International Conference on High Performance Computing, Network, Storage, and Analysis*, Denver CO USA: ACM, Nov. 2023, pp. 1697–1704. doi: 10.1145/3624062.3624250.

[48] C. Campbell *et al.*, "Superstaq: Deep Optimization of Quantum Programs." arXiv, Sep. 10, 2023. Accessed: May 06, 2024. [Online]. Available: http://arxiv.org/abs/2309.05157